\documentclass[showpacs,preprintnumbers,amsmath,amssymb,nofootinbib,superscriptaddress,twocolumn]{revtex4}

\usepackage{graphicx}
\usepackage{amsmath}
 \usepackage{color}

\begin{document}

\title{Exact results for the first-passage properties in a class of fractal networks}

\author{Junhao Peng}
\affiliation{
  School of Math and Information Science, Guangzhou University, Guangzhou 510006, China.}
\affiliation{
  Key Laboratory of Mathematics and Interdisciplinary Sciences of Guangdong Higher Education Institutes, Guangzhou University, Guangzhou 510006, China.}
   \affiliation{
  Center for Polymer Studies and Department of Physics, Boston University, Boston,
MA 02215, USA}

\author{Elena Agliari}
\affiliation{
Department of Mathematics, Sapienza Universit\`a di Roma, 00198 Rome, Italy.}
\affiliation{
Istituto Nazionale di Alta Matematica, 00198 Rome, Italy.}

\begin{abstract}
In this work we consider a class of recursively-grown fractal networks $G_n(t)$, whose topology is controlled by two integer parameters $t$ and $n$. We first analyse the structural properties of $G_n(t)$ (including fractal dimension, modularity and clustering coefficient) and then we move to its transport properties. The latter are studied in terms of first-passage quantities (including the mean trapping time, the global mean first-passage time and the Kemeny's constant) and we highlight that their asymptotic behavior is controlled by network's size and diameter. Remarkably, if we tune $n$ (or, analogously, $t$) while keeping the network size fixed, as $n$ increases ($t$ decreases) the network gets more and more clustered and modular, while its diameter is reduced, implying, ultimately, a better transport performance. The connection between this class of networks and models for polymer architectures is also discussed.
 \end{abstract}


\maketitle

\textbf{The great advances in supramolecular experimental techniques - allowing the chemical synthesis of a large variety of polymers with controlled molecular structures, including molecular fractals - make the investigation of deterministic structures timely and appealing.
In this work we introduce and analyze a class of fractal, deterministic networks
 $G_n(t)$, whose topology is controlled by two parameters $t$ and $n$.
We especially aim to highlight how transport efficiency can be influenced by the underlying topology and, to this goal, we consider a random walk process embedded  in $G_n(t)$. We first analyze  the structural properties of these networks as a function of the parameters $t$ and $n$, focusing on the fractal dimension, the modularity and the clustering coefficient. Next, we relate such topological properties to transport properties, measured in terms of the mean trapping time in the presence of a trap located at {different (sets of) nodes}, of the Kemeny's constant and of the global mean first-passage time.
Our results are all exact and provide useful hints in the design of a polymer architecture.}

\section{Introduction}
\label{intro}

Random walks constitute a fundamental model for a large number of dynamical processes, with applications ranging from physics to chemistry, biology and engineering, and even to finance, sociology and ecology (see e.g., \cite{LO93, Metzler_Klafter00, Avraham_Havlin04, Redner07, Redner09, Barkai-12,ChPe13, Metzler-2014, Agl14, Agl16, Masuda17, Barbosa_18}). Examples include light harvesting or energy  transport in polymer systems~\cite{BarKla98, Bentz03, Bentz06, BlZu81, Oshanin95}, reaction kinetics~\cite{{BenVo14, CondaBe07,BeChKl10}}, {single particle tracking experiments \cite{R3,K13,Agliari14,R2}, and search algorithms~\cite{Stanley01,Saberi04,Levin08}}. \\
An interesting quantity to look at is the  mean first-passage time (MFPT), denoted by $T_{x\rightarrow y}$, which represents the expected number of steps for a walker starting  from the site $x$ to reach the target site  $y$ for the \emph{first} time. Thus, given a system where an event is triggered as the random process reaches $y$, $T_{x\rightarrow y}$ provides the time scale for the event to occur. If  we fix the target site (also called trap) and average the MFPT over all the possible starting sites $x$, we obtain the mean trapping time (MTT) $T_y$ for trap site $y$.
In general, one can perform different kinds of average according to the physical situation to be described (see e.g., \cite{AgCasCatSar15, HaRo08, FuDo12, HvKa12, PengXu2018, MeyChe11}): being $\mathcal{G}(\Omega,E)$ the underlying (undirected) graph, with node set $\Omega$ and link set $E$, one can introduce a distribution $p_x$ (with $p_x \geq 0$ and $\sum_{x \in \Omega} p_x=1$), which represents the likelihood that $x$ is the starting node, in such a way that $T_y = \sum_{x \in \Omega} p_x T_{x\rightarrow y}$. The most common situations are the uniform one, where each site has the same probability of being selected (i.e., $p_x = 1/ |\Omega|$), and the steady-state one, where the probability that a node $x$ is selected as starting site is proportional to its degree $d_x$ (i.e., $p_x = \frac{d_x}{2|E|}$).  In this paper, we adopt the second definition of MTT, which therefore reads as\footnote{Notice that $T_{x\rightarrow x}=0$ as the walker is starting from a node $x$ that is occupied by a trap.}
\begin{eqnarray} \label{MTT_y1}
 T_y =\sum_{x\in \Omega}\frac{d_x}{2E} T_{x\rightarrow y}.
\end{eqnarray}
In inhomogeneous networks, the location of the trap $y$ may have strong effects on the MTT in such a way that $ T_y $ provides a useful indicator for the trapping efficiency of site $y$. In order to uncover the transport efficiency for the \emph{whole} network, one could further analyze the Kemeny's constant $C_{\textrm{Kemeny}}$ \cite{Aldous14, Catral10}, defined as
\begin{eqnarray} \label{KC}
C_{\textrm{Kemeny}}=\sum_{y\in \Omega}\frac{d_y}{2E}T_{x\rightarrow y},
\end{eqnarray}
and the global mean first-passage time (GMFPT) \cite{LiZh13, ZhZhGa10, ZhLi11}, referred to as  $T_{\textrm{global}}$ and defined as
\begin{eqnarray} \label{GFPT} 
 T_{\textrm{global}} =\frac{1}{N(N-1)}\sum_{x\in \Omega}\sum_{\Omega \ni y\neq x}T_{x\rightarrow y},
\end{eqnarray}
where we posed $N= |\Omega|$ (namely, $N$ is the total number of nodes in the network).

The Kemeny's constant is independent of the starting site $x$, so that it can be rewritten as
 \begin{eqnarray} \label{EKC}
C_{\textrm{Kemeny}}=\sum_{x\in \Omega}\left[ \frac{d_x}{2E}\sum_{y\in \Omega}\frac{d_y}{2E}T_{x\rightarrow y}\right].
\end{eqnarray}
The last expression highlights that the Kemeny's constant equals the average of the MFPT between any pairs of nodes while the probability that a node $u$ is selected as the starting site (target site) is $\frac{d_u}{2E}$~\cite{JIm18, Wang17, Hunter14}. On the other hand, the GMFPT is the average of the MFPT between any pairs of nodes while  all nodes of the network display the same probability of being  selected as the starting site (target site). Both the Kemeny's constant and the GMFPT are useful indicator for the transport efficiency of the network.

As mentioned above, the topology of the network has non-trivial influence on  these first-passage properties. In the past several years, considerable efforts have been devoted to evaluate the MTT and the GMFPT on  different architectures~\cite
{WuLiZhCh12, Peng14d, BeTuKo10, Weber_Klafter_10, Peng14a, Agliar08, Peng14b, ZhQiZh09, PengAgliariZhang15, ZhangXie09, MeAgBeVo12, Peng17, CoMi10, AgBu09}.

In this work we focus on a class of networks $G_n(t)$, which depend on the parameters $n,t \in \mathbb{N}$ and, as we will show, exhibit interesting topological properties such as fractality, modularity and clustering.
Also, this class of networks includes the so-called fractal cactus which has been recently shown to provide a formal representation for a kind of triangulane \cite{JurGa18,triangulane}.
Moreover, as noticed in \cite{JurGa18}, the fractal cactus can be seen as an interpolation between the Husimi cactus and the triangular Kagome lattice, which are both non-fractal.
In fact, the introduction of a deterministic class of networks with tuneable topology and transport properties is motivated by the great advances in supramolecular experimental techniques which allow the chemical synthesis of a large variety of polymers with controlled molecular structures (including molecular fractals, see e.g. \cite{Chem1,Chem2,Chem3}).
In particular, here, we first analyse the structural properties of $G_n(t)$ as a function of its parameters and we relate them to its transport properties, measured in terms of the MTT (in the presence of a trap located at a given node or at a set of given nodes), of the Kemeny's constant and of the GMFPT.

This paper is structured as follows. In Secs.~\ref{sec:Network} and \ref{sec: properities} we describe the topology of the network considered, next, in Secs.~\ref{sec:FPPs} and ~\ref{sec:FPP_Clique} we present the main method and the exact results for the first-passage properties. Finally,  Sec.~\ref{sec:Conclusion} is left for conclusions and outlooks, while technicalities on calculations are all collected in the Appendices.

\section{The network structure}
\label{sec:Network}
\begin{figure}
\begin{center}
\includegraphics[scale=0.65]{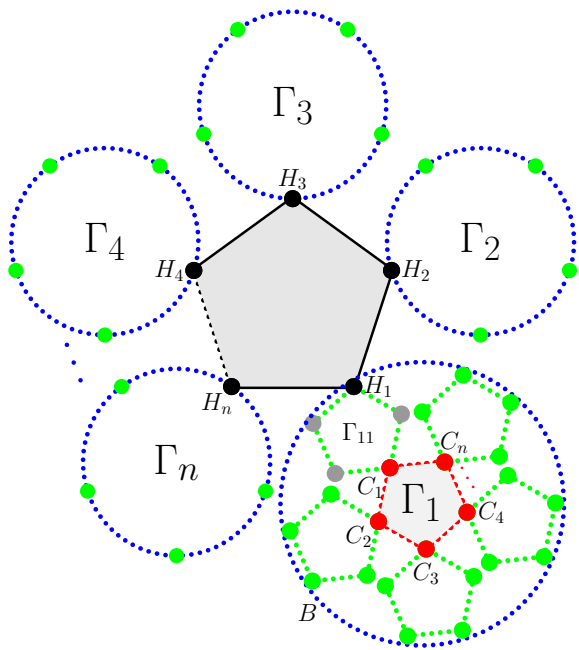}
\caption{{This figure shows a schematic representation of the network $G_n(t)$: it is composed of a central clique, represented by a gray pentagon, and $n$ subunits (here only five are depicted), referred to as $\Gamma_{i}$ ($i=1,2, ..., n$) and represented by dashed blue circles. The $n$ nodes  of the central clique, labeled as $H_{i}$ ($i=1,2, ..., n$) and colored in black, are called ``central nodes''. Nodes colored in green are some of the peripheral nodes,  which are the farthest nodes away from the central clique.
 Each subunit  $\Gamma_{i}$ is a replica of $G_n(t-1)$;  it is attached to one of the $n$ central nodes, and can be further divided into $n$ subunits. For example, $\Gamma_{1}$ is also composed of a central clique, represented by dashed red pentagon, and $n$ subunits  which are represented by dashed green pentagons. Among these, the one connected to the central clique through the node $H_1$ is labeled as $\Gamma_{11}$. The $n$ nodes  of the central clique for subunit $\Gamma_{1}$ are labeled as $C_{i}$ ($i=1,2, ..., n$) and $C_1$ is the one which is closest to $H_1$.} }
\label{fig:1}       
\end{center}
\end{figure}

\begin{figure}
\begin{center}
\includegraphics[scale=0.5]{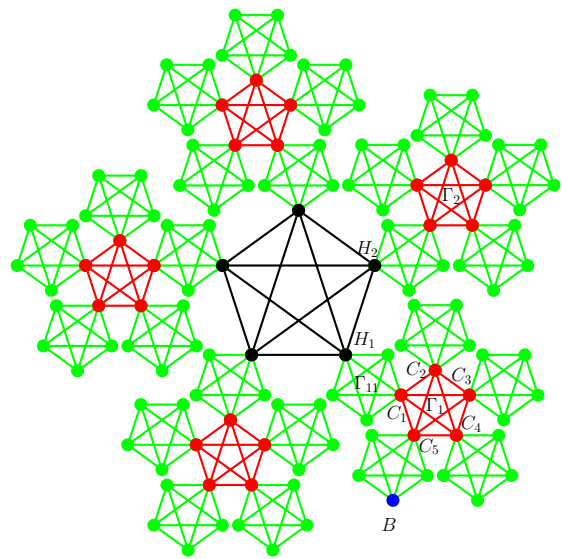}
\caption{Structure of  network $G_n(t)$ for the particular case of $n=5$ and  $t=3$; nodes of the central clique are colored with black and the node labeled as $B$ is one of the peripheral nodes, which are the farthest nodes away from  the central clique;  {the five nodes of the central clique for  subunits $\Gamma_{1}$ are labeled as $C_{i}$ ($i=1,2, ..., 5$), and $C_1$ is the node which is the closest node away from $H_1$  among the five nodes.}} 
\label{fig:2}       
\end{center}
\end{figure}

The networks considered are built in an iterative way and controlled by a positive integer parameter $n$ $(n\geq 3)$.  Let $G_n(t)$ denote the network with parameter $n$ and generation $t$ ($t \geq 1$).   The construction starts  with a  clique $K_n$   (i.e., a complete graph with $n$ nodes and $n(n-1)/2$ links), which corresponds to $G_n(1)$. For $t>1$, one starts from a  clique $K_n$ and then attaches to each of its sites a replica of $G_n(t-1)$. Fig.~\ref{fig:1} shows the structure of $G_n(t>1)$ for general $n\geq 3$ and Fig.~\ref{fig:2} shows the structure  of $G_5(3)$. The so-called fractal cactus \cite{JurGa18,triangulane} corresponds to the particular case $n=3$.
Notice that, in the following, we will highlight the dependence on $t$ only for those quantities derived recursively.
Also, we call ``central nodes'' in $G_n(t)$ the $n$ nodes of the central clique and ``peripheral nodes'' the $n(n-1)^{t-1}$ farthest nodes from the central clique. In Fig.~\ref{fig:1} and Fig.~\ref{fig:2}, nodes colored with black are the central nodes; the node labeled as $B$ is one of the peripheral nodes.

By construction, one can find that the total number of edges of $G_n(t)$ obeys the recursion relation
  \begin{equation}\label{Rec_Edges}
     E(t)=nE{(t-1)}+\frac{n(n-1)}{2}, \nonumber
  \end{equation}
 which, together with the boundary condition $E(1)= n(n-1)/2$, yields to
   \begin{equation}\label{Total_Edges}
     E(t)=\frac{n^{t+1}-n}{2}.
  \end{equation}
One can also  find that the total number of  nodes for $G_n(t)$ is
  \begin{equation}\label{Total_Nodes}
     N(t)=n^{t},
  \end{equation}
and the nodes in $G_n(t)$ can either display degree $n-1$ or $2n-2$. The total number of nodes with degree $n-1$, hereafter referred to as $N^a$, obeys the recursion relation
  \begin{equation}\label{Rec_Nta}
     N^{a}(t)=nN^{a}{(t-1)}-n, \nonumber
  \end{equation}
with the boundary condition $N^{a}(1)=n$, in such a way that
   \begin{equation}\label{Total_Nta}
     N^{a}(t)=\frac{n-2}{n-1}n^{t}+\frac{n}{n-1}.
  \end{equation}
On the other hand, the total number of nodes with degree $2n-2$, referred to as $N^b$,  is
     \begin{equation}\label{Total_Ntb}
     N^{b}(t)=N(t)- N^{a}(t)=\frac{n^{t}-n}{n-1}.
  \end{equation}

  \section{Structural properties}
  \label{sec: properities}
  In this section, we investigate several structural properties of $G_n(t)$, including the diameter, the fractal dimension, the average path length, the modularity and the clustering coefficient.

  \subsection{Diameter and fractal dimension}
   \label{subsec: diameter_F}
   We recall that the diameter of a graph is defined as the largest distance between any pair of vertices and, in the following, it will be denoted as $L_{\mathrm{max}}$.\\
By the structure sketched in Fig.~\ref{fig:1}, one can  find that the diameter of $G_n(t)$ obeys the recursion relation
   \begin{equation}\label{Rec_Diameter}
     L_{\mathrm{max}}(t)=2L_{\mathrm{max}}(t-1)+1, \nonumber
  \end{equation}
 with $L_{\mathrm{max}}(1)=1$. Thus, for any $t\geq 1$,
   \begin{equation}\label{Total_Diameter}
    L_{\mathrm{max}}(t)=2^{t}-1.
  \end{equation}
Recalling that $N(t) = n^{t}$, one has that, iteration by iteration, the total number of nodes of the network increases by a factor $n$ while the diameter increases by a factor $2$. This mirrors the change of mass in a fractal object upon the rescaling of diameter by a factor $b$:
   \begin{equation}
     N(bL_{\mathrm{max}})=b^{d_f}N(L_{\mathrm{max}}), \nonumber
  \end{equation}
where $d_f$ is the fractal dimension\footnote{Here we look upon $N(t)$ as a function of the diameter $L_{\mathrm{max}}$. {Also, it is worth mentioning that, in many applications, self-similarity emerges only statistically: in those contexts (e.g., porous media, membranes of living biological cells), percolation-like networks are widely used (see e.g., \cite{R2} for more details)}.}. In our case, $N(2L_{\mathrm{max}})=nN(L_{\mathrm{max}})$. Thus, for the networks introduced,
   \begin{equation} \label{Fractal_dimension}
    d_f=\frac{\log(n)}{\log(2)}.
  \end{equation}
For $n=3$, namely for the fractal cactus, $d_f \approx 1.585$, while for $n>3$ one has $d_f>2$.

  \subsection{Average path length}
   \label{subsec: average_path_length}
The average path length is the average of the shortest path length between any pair of nodes in the network. Let  $L_{x \rightarrow y}$ denote the shortest path length from node $x$ to node $y$ and
\begin{equation}
L_{\mathrm{total}}(t)=\sum_{x\in G_n(t)}\sum_{y\in G_n(t)}L_{x \rightarrow y},
\end{equation}
where `$x\in G_n(t)$'  means that the sum runs over all the nodes of $G_n(t)$. The  average path length can be written as
 \begin{equation}\label{Laverage}
    \overline{L}(t)=\frac{L_{\mathrm{total}}(t)}{N(t)[N(t)-1]}.
 \end{equation}
Therefore, in order to evaluate the average path length $\overline{L}(t)$,  we should first calculate $L_{\mathrm{total}}(t)$.

For $t=1$, $G_n(t)$ is a clique with $n$ nodes and the shortest path length between any two nodes is $1$. Consequently, $L_{\mathrm{total}}(1)=n(n-1)$. Exploiting the equivalence of the $n$ subunits $\Gamma_{i}$ ($i=1,2, ..., n$), 
for $t>1$, we find
  \begin{eqnarray}\label{Rec_LTotal}
     L_{\mathrm{total}}(t)
    &=&nL_{\mathrm{total}}(t-1)\nonumber\\
    & &+n^{2t-2}(n-1)[(n-1) 2^t-n+2],
 \end{eqnarray}
as derived in details in Appendix ~\ref{Sec:Rec_LTotal}. 
 Using Eq.~(\ref{Rec_LTotal}) recursively, we obtain
  \begin{eqnarray}\label{LTotal}
 L_{\mathrm{total}}(t) &=&
   n^{t} \Big[4(n-1)^2 \frac{(2n)^{t-1}-1}{2n-1} \nonumber \\
   &-&
    (n-2) n^{t-1} +2n-3 \Big].
 \end{eqnarray}
Replacing $N(t)$ with $n^t$ and plugging Eq.~(\ref{LTotal}) into Eq.~(\ref{Laverage}), we get
  \begin{equation} \label{Average_pathlength}
   \overline{L}(t)=\frac{n^{t-1}[2^t (n-1)^2 - (2n-1)(n-2)] -1 }{(2n-1)(n^t-1)}.
  \end{equation}

Therefore, the average path length $\overline{L}(t)\approx 2^t\sim [N(t)]^{\frac{\log(2)}{\log(n)}}$.
In ``small world'' networks, the typical distance between two randomly chosen nodes scales logarithmically with the number of nodes and, in the current case, this means a linear scaling with $t$. This condition is not fulfilled by $\overline{L}(t)$, however, it should be noted that the distribution of the distances $L_{x \rightarrow y}$ is rather heterogenous.

\subsection{Modularity}
\label{Sec: Modularity}

The modularity measures the strength of division of a network into modules~\cite{Newman2004, Newman2006}: networks with high modularity display dense connections between the nodes belonging to the same module but sparse connections between nodes in different modules.
A possible definition of modularity is the following: fix a certain division of the network sites into a certain set of modules and, for each pair $(i,j)$ of modules, evaluate the fraction $e_{ij}$ of the edges connecting one site in module $i$ and one site in module $j$; evaluate the fraction $a_i$ of ends of edges that are attached to vertices in module $i$; then, the modularity $Q$ is given by $Q=\sum_i (e_{ii} - a_i^2) \in [-1,1]$, where the sum runs over all modules. In this way, a high modularity means that the number of edges within the module is larger than that you expect by chance.

The network considered here is clearly composed of $n$ modules (i.e., $\Gamma_{i}$, $i=1,2, ..., n$) as shown in Fig.~\ref{fig:1} and, according to such a division, we can evaluate the $n\times n$ symmetric matrix $e$ whose elements are given by
 \begin{equation}
 e_{ij}=\left\{\begin{array}{ll} \frac{E(t-1)}{E(t)} & j=i\\ \frac{1}{2E(t)}  & j\neq i \end{array}\right.\nonumber.
 \end{equation}
Then, the fraction of edges that connect to nodes in module $\Gamma_{i}$ is
 \begin{equation}
 a_{i}=\sum_{j=1}^n e_{ij}=\frac{2E(t-1)+n-1}{2E(t)}=\frac{1}{n}\nonumber.
 \end{equation}
Exploiting the previous results we get that the modularity for $G_n(t)$ is
 \begin{equation}\label{Modularity}
 Q=\sum_{i=1}^n \left[\frac{E(t-1)}{E(t)}  -\frac{1}{n^2} \right]=\frac{n^{t+1}-n^t-n^2+1}{n^{t+1}-n},
 \end{equation}
and, in the limit of large size (i.e., $t\rightarrow \infty$),  $Q\rightarrow \frac{n-1}{n}$. \\
Finally, we notice that, given two graphs $G_{n_1}(t_1)$ and $G_{n_2}(t_2)$ with (approximately) the same size (i.e., $n_1^{t_1} \approx n_2^{t_2}$) with $n_1 < n_2$ to fix ideas, then $Q_1 < Q_2$, as long as $N>n_1 n_2+1$, which, for $t_2>2$, is always fulfilled.

\subsection{Clustering coefficient}
\label{Sec: Clustering_coefficient}
Here, we calculate the local clustering coefficient for an arbitrary node and the average clustering coefficient for the  whole network.
The local clustering coefficient of a given node is the ratio between the total number of edges that
actually exist between its $k$ nearest neighbours and the potential number of edges $k(k-1)/2$
between them. The clustering coefficient of the whole network is obtained by averaging the local clustering coefficient over all
its nodes~\cite{Watts_St1998, Albert_Barab2002}.

As explained in Sec.~\ref{sec:Network}, the nodes of $G_n(t)$ can either have degree $n-1$ (type $a$) or degree $2n-2$ (type $b$).
For a node with degree $n-1$, there is an edge between any pairs of its nearest neighbours. Thus, the local clustering coefficient for this kind of nodes is $$C^a=1.$$
For a node with degree $2n-2$, which is the intersection of two cliques, there are $(n-1)(n-2)$ edges
between  its nearest neighbours, then the local clustering coefficient for this kind of nodes is $$C^b=\frac{2(n-1)(n-2)}{(2n-2)(2n-3)}=\frac{n-2}{2n-3}.$$
As a consequence, the average clustering coefficient for the whole network is
 \begin{equation}\label{A_Modularity}
 \overline{C}=\frac{N^a(t) C^a+N^b(t) C^b}{N(t)}=\frac{2n-4}{2n-3}+\frac{n^{1-t}}{2n-3},
 \end{equation}
and, in the limit of large size (i.e., $t\rightarrow \infty$),  $\overline{C}\rightarrow \frac{2n-4}{2n-3}$.
As expected, the network is therefore highly clustered.
Also, taking two graphs $G_{n_1}(t_1)$ and $G_{n_2}(t_2)$ as in the previous subsection, we get that $ \overline{C}_1 <  \overline{C}_2$ for any $n_1 < n_2$.

\section{First-passage to a node }
\label{sec:FPPs}

{In this section, we analyze and obtain exact results for the MTT (see Eq.~(\ref{MTT_y1})), the GMFPT (see Eq.~(\ref{GFPT})) and the Kemeny's constant (see Eq.~(\ref{EKC})) of the network $G_n(t)$. More precisely, the MTT is evaluated rigorously in the presence of a trap located at one of the central nodes and in the presence of a trap located at one of the peripheral nodes; also, we find bounds for the MTT when the trap is set on an arbitrary node and this allows us to get the related scaling behavior with respect to the network size}.
 The method presented is based on the connection between the MFPT and the effective resistor. Here we just present the main idea and the main results, while he detailed derivations are collected in the Appendices \ref{sec: Rel_R_L}-\ref{sec: eTSUBcenter}.

\subsection{Main method}
\label{sec:gen_meth}

{The commute time $T_{x \leftrightarrow y}$ between a couple of arbitrary nodes $x$ and $y$ in $G_n(t)$ is defined as}
$$T_{x \leftrightarrow y}= T_{x\rightarrow y}+ T_{y\rightarrow x}$$  and the MFPT from node $x$ to $y$ can be expressed in terms  of commute times as~\cite{Te91}
\begin{equation}
T_{x\rightarrow y}=\frac{1}{2}\left[T_{x \leftrightarrow y}+\sum_{u\in G_n(t)}\frac{d_u}{2E(t)}(T_{y \leftrightarrow u}-T_{x \leftrightarrow u} ) \right],
\label{FXY}
\end{equation}
where the sum runs over all the nodes of $G_n(t)$. Now, the commute time can be estimated straightforwardly exploiting the reciprocity theorem of electrical networks, properly restated in terms of random walks \cite{DoyleSnell,foster,Te91}. In fact, let us look at the graph $G_n(t)$ as an electrical network, where each edge corresponds to a unit resistor, and let us denote with $\mathfrak{R}_{x\rightarrow y}$ the effective resistance  between two nodes $x$ and $y$, then one can prove that \cite{Te91}
\begin{equation}
T_{x\leftrightarrow y}=2E(t)\mathfrak{R}_{x\rightarrow y},
\label{KR}
\end{equation}
where $E(t)$ is the total number of edges of $G_n(t)$.  For the networks considered here, we find 
 \begin{equation}\label{Rel_R_L}
 \mathfrak{R}_{x\rightarrow y}=\frac{2}{n}L_{x \rightarrow y},
 \end{equation}
 where $L_{x \rightarrow y}$ denotes  the shortest path lengths from node $x$ to node $y$. The detailed proof of Eq.~(\ref{Rel_R_L}) is presented in Appendix~\ref{sec: Rel_R_L}. Combining Eq. (\ref{KR}) and Eq. (\ref{Rel_R_L}) we get
 \begin{equation}
T_{x \leftrightarrow y}=\frac{4}{n}E(t)L_{x  \rightarrow y}.
\label{KL}
\end{equation}
The GMFPT of the network $G_n(t)$ can be written as
\begin{eqnarray} \label{GMFPT}
T_{\textrm{global}}
                &=&\frac{1}{2}\frac{1}{N(t)(N(t)-1)}\sum_{x\in G_n(t)}\sum_{y\neq x}( T_{x\rightarrow y}+T_{y\rightarrow x})\nonumber \\
                 &=&\frac{1}{N(t)(N(t)-1)}\sum_{x\in G_n(t)}\sum_{y\neq x}\frac{2}{n}E(t)L_{x \rightarrow y}\nonumber \\
                 &=&\frac{2E(t)}{nN(t)(N(t)-1)}L_{\mathrm{total}}(t)\nonumber \\
                 &=&\frac{L_{\mathrm{total}}(t)}{n^t},
\end{eqnarray}
where
\begin{equation}
L_{\mathrm{total}}(t)=\sum_{x\in G_n(t)}\sum_{y\in G_n(t)}L_{x \rightarrow y}
\end{equation}
is the sum of shortest path length between any pairs of nodes of $G_n(t)$.

Substituting $T_{x \leftrightarrow y}$ with the right-hand side of Eq.~(\ref{KL}) in Eq.~(\ref{FXY}),
$T_{x \rightarrow y}$ can be rewritten as
\begin{eqnarray} \label{EMTTY}
T_{x \rightarrow y}
      &=&\frac{2E(t)}{n}[{L_{x \rightarrow y}}+W_y(t)-W_x(t)],
\label{FXYL}
\end{eqnarray}
where
\begin{equation}
W_v(t)=\sum_{u \in G_n(t)}\frac{d_u}{2E(t)}{L_{u \rightarrow v}}.
\label{WY}
\end{equation}
Replacing $T_{x \rightarrow y}$ with the right-hand side of Eq.~(\ref{EMTTY}) in {Eq.~(\ref{MTT_y1})} and (\ref{EKC}), we get, respectively, the MTT for $y$ 
\begin{eqnarray} \label{MTT_y}
T_y(t)
&=&\frac{2}{n}E(t)[2W_y(t)-\Sigma(t)],
\end{eqnarray}
and the Kemeny's constant
\begin{eqnarray} \label{KC}
C_{\textrm{Kemeny}}&=&\frac{2E(t)}{n}\sum_{y\in G_n(t)}\frac{d_y}{2E(t)}\left[L_{xy}+W_y(t)-W_x(t)\right]\nonumber\\
    &=&\frac{2E(t)}{n}\sum_{y\in G_n(t)}\frac{d_y}{2E(t)}W_y(t)\nonumber\\
    &=&\frac{2E(t)}{n}\Sigma(t),
\end{eqnarray}
where
\begin{eqnarray}
\nonumber
\Sigma(t)&=&\sum_{y\in G_n(t)}\frac{d_y}{2E(t)}W_y\\
\nonumber
&=&\sum_{x\in G_n(t)}\sum_{y\in G_n(t)}\frac{d_x}{2E(t)}\frac{d_y}{2E(t)} {L_{x \rightarrow y}}.
\end{eqnarray}

Therefore, the global quantities $T_{\textrm{global}}$ and $C_{\textrm{Kemeny}}$ are expressed in terms of $L_{\textrm{total}}(t)$ and $\Sigma(t)$, respectively; also, in order to evaluate $T_y$ for a given $y$, we first need to obtain $W_y(t)$.

\subsection{Main results }
\label{sec:Results}

Here we exploit the results obtained in the previous subsection to derive the explicit expression of $T_{\textrm{global}}$, $C_{\textrm{Kemeny}}$ and $T_y$, with $y$ being either a central node or a peripheral node; {an estimate for $T_y$, with $y$ being an arbitrary node is also provided}.

Inserting the result of $L_{\textrm{total}}(t)$ as shown in Eq.~(\ref{LTotal}) into Eq.~(\ref{GMFPT}), 
we obtain the GMFPT of $G_n(t)$ as
  \begin{eqnarray}\label{GMFPT_ER} 
  \nonumber
  T_{\textrm{global}}&=& (2n-3)2^{t-1}n^{t-1}-n^{t-1}(n-2)+\frac{2^{t-1}n^{t-1}-1}{2n-1}\\
  &\approx& (2n)^{t} \frac{2 (n-1)^2}{n(2n-1)},
 \end{eqnarray}
 where, in the last line, we highlighted the leading term for large $n$ and $t$.
\newline
 Moreover, as derived in Appendix~\ref{Sec:Sigma_t},
 \begin{eqnarray}\label{EQ_Sigma}
    \Sigma(t)&=&\frac{n^{t}[2^{t}n^{t}(2n+\frac{1}{2n}-3)-2n^{t+1}+5]-2}{2n(n^t-1)^2}\nonumber \\
    & &+\frac{n^{t-1}(2^{t-1}n^{t-1}-1)}{2(2n - 1)(n^t-1)^2},
 \end{eqnarray}
and, plugging this expression into Eq.~(\ref{KC}), we obtain the Kemeny's constant: 
 \begin{eqnarray}\label{Kemeny_ER}
    C_{\mathrm{Kemeny}}&=&\frac{2^{t+1}n^{2t}(n-1)^2+3n + n^{2t+1} - 2n^{2t+2} - 2}{n(2n - 1)(n^{t} - 1)}\nonumber \\
  & &+\frac{5n - 3}{n(2n - 1)} \approx (2n)^t \frac{2(n-1)^2}{n(2n-1)}.
 \end{eqnarray}

Analyzing the shortest path length from an arbitrary node to a central node $H_1$ and to a peripheral node $B$ (see  Fig.~\ref{fig:1}), and calculating their weighted mean, we obtain
\begin{eqnarray}
W_B(t)&=&\sum_{u \in G_n(t)}\frac{d_u}{2E(t)}{L_{u \rightarrow B}}\nonumber\\
    &=&2^t-\frac{n+1}{n}-\frac{(n^{t-1}-1)(2^t-1)}{(n^t - 1)}.
\label{WB}
\end{eqnarray}
and
\begin{eqnarray}
W_{H_1}(t)&=&\sum_{u \in G_n(t)}\frac{d_u}{2E(t)}{L_{u \rightarrow H_1}}\nonumber\\
&=&2^{t-1} -\frac{ 1}{n} -\frac{2^{t-1} (n^{t-1} - 1)}{(n^t - 1)}.
\label{WH}
\end{eqnarray}
The detailed derivation of Eqs.~(\ref{WB}) and (\ref{WH}) is presented in Appendix~\ref{Sec:WB} and ~\ref{Sec:WH}, respectively.

Inserting Eqs.~(\ref{EQ_Sigma}) and (\ref{WB}) into Eq.~(\ref{MTT_y}), we obtain  the MTT while a trap is located at  a peripheral node $B$, that is
 \begin{eqnarray}\label{MTTB}
   T_B(t)&=&\frac{ ( 2^{t+1}-2)( n^{2t+2}-2n^{t+2} +2n-1)}{n(2n - 1)(n^{t} - 1)}\\
   \nonumber
      & -&\frac{ ( 2^{t+1}-1)(n^{t+1}-2n+1)}{n(2n - 1)} \approx (2n)^t \frac{2(n-1)}{(2n-1)}.
 \end{eqnarray}
Similarly,  the MTT while a trap is located at central node $H_1$ is
 \begin{eqnarray}\label{MTTH}
  T_{H_1}(t)&=&\frac{2^{t}n^{2t}(n-1)+n^{2t}(2n^2-5n+2)}{n(2n - 1)(n^{t} - 1)}  \\
  \nonumber
  &-&   \frac{2^{t}n^{t}(2n^2-3n+1) - n^{t}(3n-1) + 2n-1 }{n(2n - 1)(n^{t} - 1)}\\
  \nonumber
  &\approx& (2n)^t \frac{(n-1)}{n(2n-1)}.
 \end{eqnarray}

{We notice that $T_{\mathrm{global}}$ and $C_{\mathrm{Kemeny}}$ display the same leading term as $n$ and $t$ are large, while $T_B / T_{H_1} \approx n/2 >1$, as expected.
However, in the limit of large size (i.e., as $t\rightarrow \infty$) the scaling behavior for all the quantities considered is the same and given by }
$(2n)^t$.
Otherwise stated, recalling that $N(t)=n^t$ and that $L_{\textrm{max}}(t) = 2^t-1$, we get
\begin{eqnarray} \label{Scalings}
T_{\mathrm{global}} \sim C_{\mathrm{Kemeny}}\sim  T_B \sim  T_{H_1} \sim L_{\textrm{max}}(t) \times N(t)  \label{eq:asym}.
\end{eqnarray}

{The same scaling also holds for the MTT when the trap i located at an arbitrary node $y \in G_n(t)$. In fact, as proved in Appendix~\ref{Sec:Comp},
  \begin{equation}\label{Max_Min_Wy}
 W_{H_1}(t)\leq W_{y}(t) \leq W_{B}(t),
 \end{equation}
and, recalling Eq.~(\ref{EMTTY}), we get,
   \begin{equation}\label{Max_Min_Ty}
 T_{H_1}(t)\leq T_{y}(t) \leq T_{B}(t),
 \end{equation}
that is, $\min_{y \in G_n(t)} T_y =  T_{H_1}(t)$ and $\max_{y \in G_n(t)} T_y =  T_{B}(t)$,
thus
 $$T_{y} \sim L_{\textrm{max}}(t) \times N(t)$$
 for any node $y$.
 }
{We check this result by considering the MTT for the node $C_1$ (see Fig.~\ref{fig:1}): as derived in Appendix~\ref{sec: eTSUBcenter},
 \begin{eqnarray}\label{WC_WH}
  W_{C_1}(t)-W_{H_1}(t)=(2^{t-2}-1)\frac{n^{t-2}(n-1)^2}{n^t-1},
 \end{eqnarray}
 which yields to
 \begin{eqnarray}\label{MTTC1}
  T_{C_1}(t)&=&\frac{2}{n}E(t)[2W_{C_1}(t)-\Sigma(t)]\nonumber \\
  &=&T_{H_1}(t)+\frac{2}{n}E(t)[2W_{C_1}(t)-2W_{H_1}(t)]\nonumber \\
  &=&\frac{2^{t}n^{2t}(n-1)-2^{t}n^{t}(2n^2-3n+1)-2n+1 }{n(2n - 1)(n^{t} - 1)} \nonumber \\
  & &+\frac{n^{2t}(2n^2-5n+2)+n^{t}(3n-1)}{n(2n - 1)(n^{t} - 1)}\nonumber \\
  & &+2(2^{t-2}-1)n^{t-2}(n-1)^2 \nonumber \\
  &\approx&(2n)^t \frac{(2n^2-n+1)(n-1)}{2n^2(2n-1)}.
 \end{eqnarray}
 }

\begin{figure}
\begin{center}
\includegraphics[scale=0.25]{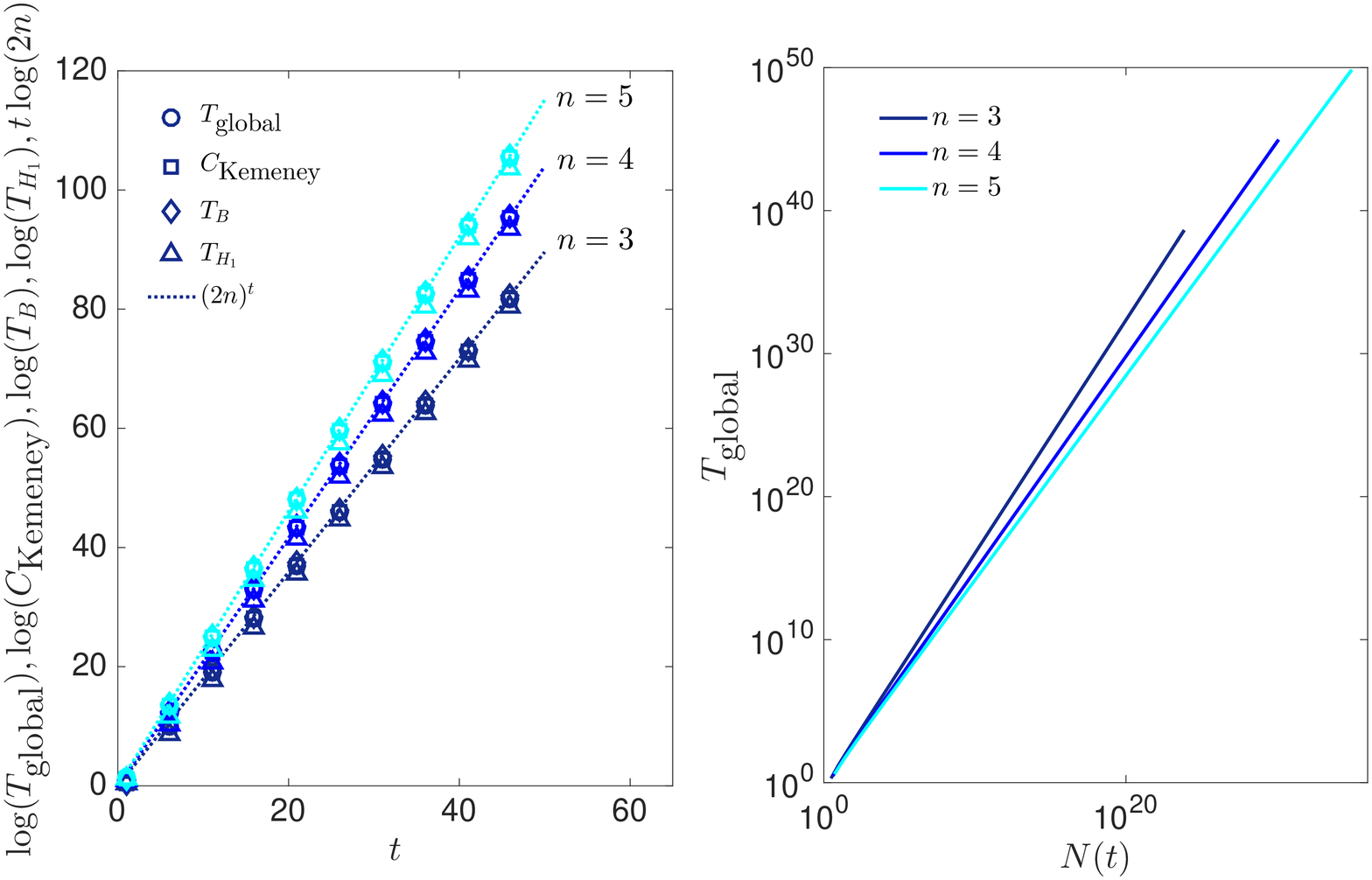}
\caption{Left panel: the quantities evaluated in this subsection (represented in different symbols as explained by the legend) are shown as a function of the graph generation $t$ and for different choices of $n$, in a log-$y$ scale. Notice that the asymptotic behavior (dashed line) highlighted in Eq.~(\ref{eq:asym}) is nicely recovered even for relatively small values of $t$. Right panel: the GMFPT is shown versus the system size $N(t)$ in a log-log scale and for different choices of $n$, as explained by the legend. Notice that, for a given size, increasing $n$ implies a better performance. The other first-passage quantities considered in this work display similar behaviors.}
\label{fig:3}       
\end{center}
\end{figure}

The scaling behaviors highlighted in this subsection allows us to state that, for two networks $G_{n_1}(t_1)$ and $G_{n_2}(t_2)$ displaying approximately the same size, namely $n_1^{t_1} \approx n_2^{t_2}$, with $n_1 < n_2$ to fix ideas, the quantities $T_{\mathrm{global}}, C_{\mathrm{Kemeny}}$, {and $T_{y}$ (with $y$ an arbitrary node)} related to the former are exponentially larger since $t_1>t_2$. Thus, although a larger $n$ makes the walker more likely to ``get lost'' in each single module, and although a larger $n$ improves the modularity and the clustering (as shown in Secs.~\ref{Sec: Modularity} and \ref{Sec: Clustering_coefficient}, respectively), for a given size, the transport efficiency of $G_n(t)$ is ultimately controlled by its diameter; this is highlighted by the plots in Fig.~\ref{fig:3}.

\section{First-passage to a clique }
\label{sec:FPP_Clique}

\begin{figure}
\begin{center}
\includegraphics[scale=0.25]{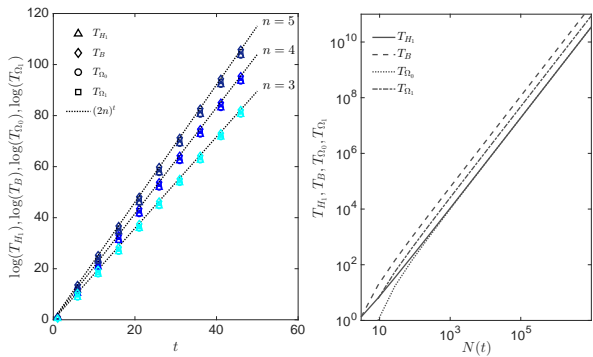}
\caption{{Left panel: the MTTs evaluated in this work (represented in different symbols as explained by the legend) are shown as a function of the graph generation $t$ and for different choices of $n$, in a log-$y$ scale. Notice that the asymptotic behavior (dashed line) highlighted in Eqs.~(\ref{eq:asym}) and (\ref{eq:asym2}) is nicely recovered even for relatively small values of $t$. Right panel: the same quantities are shown versus the system size $N(t)$ in a log-log scale and for $n=3$ (analogous behaviors are found for different choices of $n$). Notice that, for a given size, the largest MTT corresponds to the case where the trap is located at a single peripheral node, while when the trap is located at the central node or at the central clique the related MTTs appear overlapped.}}
\label{fig:4}       
\end{center}
\end{figure}

{In this section we consider the case of an extended trap, namely the case where the trap involves a set of nodes $\Omega' \subset \Omega$. Then, the trapping time $T_{x \rightarrow \Omega'}$ is given by the time taken by the walker starting from $x$ to first reach any node labelled as trap and the MTT, denoted with $T_{\Omega'}$, shall be obtained by averaging over all possible starting nodes in analogy to Eq.~(\ref{MTT_y1}), that is
\begin{equation}
T_{\Omega'} = \sum_{x \in \Omega} \frac{d_x}{2E} T_{x \rightarrow \Omega'}.
\end{equation}
Here we will especially focus on the case where $\Omega'$ is a clique. In this case, the MTT can be expressed in terms of the MTT for a given node and the MFPT between two given nodes, which can be  exactly calculated by using the method presented in Sec. \ref{sec:FPPs}.
For instance, let us consider as trap the central clique of $G_n(t)$, referred to as $\Omega_0$, and the central clique of one of the subunit $\Gamma_i$ ($i=1,2, ..., n$), referred to as $\Omega_1$; in any case the number of trapping nodes is given by $n$ independently of the generation $t$.}
\newline
{
For the former the MTT  can be expressed as
\begin{eqnarray} \label{Tcenter}
T_{\Omega_0}(t)&=&\sum_{x\in G_n(t)}\frac{d_x}{2E(t)}T_{x\rightarrow \Omega_0}\nonumber
 = \sum_{i=1}^n\sum_{x\in \Gamma_i}\frac{d_x}{2E(t)}T_{x\rightarrow \Omega_0}\nonumber\\
    &=&\sum_{i=1}^n\sum_{x\in \Gamma_i}\frac{d_x}{2E(t)}T_{x\rightarrow H_i}\nonumber\\
    &=&n\frac{E(t-1)}{E(t)}\sum_{x\in \Gamma_1}\frac{d_x}{2E(t-1)}T_{x\rightarrow H_1}\nonumber\\
    &=&\frac{n^t-n}{n^t-1}T_B(t-1),
\end{eqnarray}
and replacing $T_B(t-1)$ from  Eq.~(\ref{MTTB}) in Eq~(\ref{Tcenter}), we get 
\begin{eqnarray} \label{eTcenter}
T_{\Omega_0}(t)
       &=&\frac{ ( 2^{t}-2)( n^{2t}-2n^{t+1} +2n-1)}{(2n - 1)(n^{t} - 1)}\nonumber\\
      & &-\frac{ (n^{t-1} - 1)( 2^{t}-1)(n^{t}-2n+1)}{(2n - 1)(n^{t} - 1)} \nonumber \\
      &\approx& (2n)^t \frac{n-1}{n(2n-1)},
\end{eqnarray}
where, in the last passage, we highlighted the leading term for large $n$ and $t$.}
\newline
{
For the latter, without loss of generality, let the central clique of  subunit $\Gamma_1$  be trap,  label the $n$ nodes of the central clique for $\Gamma_1$ as $C_i$ ($i=1,2, ..., n$) (see Fig.~\ref{fig:1} for the general  case and Fig.~\ref{fig:2} for the particular case of $n=5$ and $t=3$), and let $\Omega_1=\{C_1, C_2, \cdots, C_n\}$. Then, the MTT 
can be expressed as
\begin{eqnarray} \label{TSUBcenter}
&&T_{\Omega_1}(t) =\sum_{x\in G_n(t)}\frac{d_x}{2E(t)}T_{x\rightarrow \Omega_1}\nonumber\\
    &=&\sum_{x\in \Gamma_1}\frac{d_x}{2E(t)}T_{x\rightarrow \Omega_1}+\sum_{i=2}^n\sum_{x\in \Gamma_i}\frac{d_x}{2E(t)}T_{x\rightarrow C_1}\nonumber\\
    &=&\frac{E(t-1)}{E(t)}\sum_{x\in \Gamma_1}\frac{d_x}{2E(t-1)}T_{x\rightarrow \Omega_1}\nonumber\\
    &&+(n-1)\sum_{x\in \Gamma_2}\frac{d_x}{2E(t)}[T_{x\rightarrow H_2}+T_{H_2\rightarrow H_1}+T_{H_1\rightarrow C_1}]\nonumber\\
    &=&\frac{E(t-1)}{E(t)}T_{\Omega_0}(t-1)+\frac{n-1}{2E(t)}T_{H_1\rightarrow C_1}\nonumber\\
    & & +(n-1)\frac{E(t-1)}{E(t)}T_B(t-1)\nonumber\\
    &&+\frac{n-1}{n}[T_{H_2\rightarrow H_1}+T_{H_1\rightarrow C_1}],
\end{eqnarray}
%
%
and replacing $T_{\Omega_0}(t-1)$ and $T_B(t-1)$ from Eqs.~(\ref{eTcenter}) and (\ref{MTTB}) in Eq.~(\ref{TSUBcenter}), calculating $T_{H_2\rightarrow H_1}$ and $T_{H_1\rightarrow C_1}$ and inserting them into Eq.~(\ref{TSUBcenter}), we finally obtain (see Appendix~\ref{sec: eTSUBcenter} for a detailed derivation)
\begin{eqnarray} \label{eTSUBcenter}
&&T_{\Omega_1}(t)\nonumber\\
    &=&(2n)^{t-1}(n-1) \left[ 1 + \frac{1}{2n^2(2n-1)} \right]-n^{t-2}(n^2-2n+2) \nonumber\\
      & &-\frac{(2n)^{t-2} (n-1)^2(6n^2 - n + 1) }{n(2n - 1)(n^{t} - 1)}\nonumber\\ %
      & &+\frac{ 5n^t -6n^{t-1}+2n^{t-2}-2n +1 }{(2n - 1)(n^{t} - 1)} \nonumber\\
    &\approx&(2n)^t\left[  \frac{ n-1}{2n}+\frac{ n-1}{8n^4-4n^3}\right].
\end{eqnarray}
In the limit of large size (i.e., as $t\rightarrow \infty$), we get the same scaling found for the quantities analysed in the previous section, that is
\begin{eqnarray}
T_{\Omega_0}(t) \sim T_{\Omega_1}(t) \sim (2 n)^{t} \sim L_{\textrm{max}}(t) \times N(t).   \label{eq:asym2}
\end{eqnarray}
Remarkably, the size of the trapping-node set does not yield to qualitative effects on the large-size scaling and, in the case of central location ($T_{H_1}$ and $T_{\Omega_0}$), even the leading terms are the same;
this is highlighted by the plot in Fig.~\ref{fig:4}.}

\section{Conclusion} \label{sec:Conclusion}

In this work we considered a class of recursively grown networks, referred to as $G_n(t)$, whose topology is controlled by two parameters $t, n \in \mathbb{N}$, with $n >2$ and $t$ being the network generation. This kind of networks provides an interesting model for polymer structures and, in particular, the case $n=3$, also known as fractal cactus, constitutes the formal representation of the branched triangulane which has been experimentally synthesized up to the generation $t=3$ \cite{JurGa18,triangulane}.
\newline
First, we analyzed the topology of $G_n(t)$ showing that it exhibits a fractal nature, with fractal dimension $d_f = \log(n)/\log(2)$, a diameter $L_{\textrm{max}}=2^t-1$, a modularity $Q \rightarrow (n-1)/n$, and an average clustering coefficient $\overline{C} \rightarrow (2n-4)/(2n-3)$, where the limits are taken with respect to the system size (i.e., as $t \rightarrow \infty$). Remarkably, although the network is highly clustered, its diameter grows relatively fast with the system size so that $G_n(t)$ is not ``small-world''. We also showed that, for a fixed network size $N=n^t$, the modularity and the average clustering are mainly ruled by $n$, namely a larger $n$ implies a larger $Q$ and $\overline{C}$.
\newline
Next, we addressed the calculation of mean first-passage quantities in order to get information about the efficiency of transport on $G_n(t)$, as $t$ and $n$ are tuned. In particular, we focused on the GMFPT, 
on the Kemeny's constant, 
and on the MTT when the trap is located on a peripheral or on a central node, 
{or when the trap involves a set of nodes making up a clique. In the limit of large generation} all these quantities display the same scaling behaviour given by $ L_{\textrm{max}} \times N \sim (2n)^t$. Therefore, in these structures, the target position does not qualitatively influence the efficiency of the searching process. Also, fixed a size $N$, the time scale for such first-passage quantities is ruled mainly by $t$, on the contrary of the modularity and of the average clustering coefficient.

{The results obtained in this work provide interesting hints for the design of a polymer (or of a generic architecture) embedding a diffusion process.
First, we stress that the fractal dimension of the cactus is below $2$, while other $n$ lead to arbitrarily large values of $d_f$. This implies a transition from recurrent to transient diffusion processes which may have remarkable effects even in (large) finite systems: when considering diffusion on polymers in the presence of a reaction center, when $n$ is large the walker could get lost for long time before eventually reaching the target.
Moreover, the target location does not qualitatively influence the large-size limit of the MTT, in such a way that if one aims to control the reaction time, adjusting the trap position only allows for a fine tuning. Even adjusting the extent of the trap would not be qualitatively effective. Thus, in order to shorten the reaction time, with the number of traps meant as a cost, the optimal solution is setting a single, central trap.}

{Finally, we mention a few extensions that may mimic interesting applications and that would be worth of investigation. One could consider the case where the walker can perform long-range jump (see e.g., \cite{R1}), the case where the trap is as well allowed to move (see e.g., \cite{ACCS-15}), and the case of more general reactions, like autocatalytic, coalescence, annihilation, etc. (see e.g., \cite{Avraham_Havlin04}).}

\acknowledgments
{
\noindent
The authors are grateful to Prof. Shlomo Halvin for  enlightening discussions and acknowledge the National Natural Science Foundation of China (Grant No. 61873069).
The Guangzhou University School of Math and Information Science is supported by the National Key R\&D Program of China (Grant No. 2018YFB0803604), by the special innovation project of colleges and universities in  Guangdong (Grant No. 2017KTSCX140), and by he National Natural Science Foundation of China (Grant No. 61772147). The Boston  University Center for Polymer Studies is supported by NSF Grants  PHY-1505000, CMMI-1125290, and CHE-1213217, and by DTRA Grant  HDTRA1-14-1-0017.\\
EA is grateful to GNFM-INdAM (Progetto Giovani 2018) and to Sapienza University of Rome (Progetto Ateneo RG11715C7CC31E3D and RM116154CD9961A3) for financial support.  \\

}

\appendix

 \section{Derivation of Eq.~(\ref{Rec_LTotal})}
\label{Sec:Rec_LTotal}
In this Appendix, we will derive the  recursion relation of $L_{\mathrm{total}}(t)$ as shown in Eq.~(\ref{Rec_LTotal}).
 Before preceding, we should  calculate the sum of the shortest path length from arbitrary node to a peripheral node (e.g. node $B$ as shown in Fig.~\ref{fig:1}), defined as
\begin{equation}
L_B(t)=\sum_{x\in G_n(t)}L_{x \rightarrow B}
\end{equation}

 It is easy to obtain that $L_B(1)=n-1$. For $t>1$, as shown in Fig.~\ref{fig:1},  $G_n(t)$ is  composed of a central clique and $n$ subunits $\Gamma_{i}$ ($i=1,2, ..., n$).  Each subunit  $\Gamma_{i}$ is a replica of $G_n(t-1)$ and it is attached to one of the $n$ central nodes. Thus
 \begin{eqnarray}\label{REC_S_B}
   L_B(t)&=&\sum_{x\in \Gamma_{1}}L_{x \rightarrow B} +\sum_{i=2}^n\sum_{x\in \Gamma_{i}}L_{x \rightarrow B} \nonumber \\
    &=&L_B(t-1)+(n-1)\sum_{x\in \Gamma_{2}}(L_{x \rightarrow H_2}+1+L_{B \rightarrow H_1})\nonumber \\
    &=&L_B(t-1)+(n-1)\sum_{x\in \Gamma_{2}}L_{x \rightarrow H_2}+(n-1)n^{t-1}2^{t-1}\nonumber\\
    &=&nL_B(t-1)+(n-1)n^{t-1}2^{t-1}.
 \end{eqnarray}
 Using Eq.~(\ref{REC_S_B}) recursively, we get
 \begin{eqnarray}\label{ER_S_B}
    L_B(t)&=&n^2_B(t-2)+(n-1)n^{t-1}[2^{t-2}+2^{t-1}]\nonumber\\
    &=&\cdots \nonumber\\
    &=&n^{t-1}_B(1)+(n-1)n^{t-1}[2+2^2+\cdots+2^{t-1}] \nonumber\\
    &=&(n-1)n^{t-1}[1+2+2^2+\cdots+2^{t-2}+2^{t-1}] \nonumber\\
     &=&(n-1)n^{t-1}(2^{t}-1).
 \end{eqnarray}

%
Now, we derive the  recursion relation of $L_{\mathrm{total}}(t)$ as shown in Eq.~(\ref{Rec_LTotal}). For $t=1$, $G_n(t)$ is a clique with $n$ nodes. The shortest path length between any two nodes is $1$. Therefore $L_{\mathrm{total}}(1)=n(n-1)$. Note the equivalence of the $n$ subunits $\Gamma_{i}$ ($i=1,2, ..., n$) as shown in Fig.~\ref{fig:1}, for $t>1$,
  \begin{eqnarray}
    && L_{\mathrm{total}}(t)\nonumber\\
    &=&n\sum_{x\in  \Gamma_{1}}\sum_{y \in G_n(t)}L_{x \rightarrow y}\nonumber\\
    &=&n\left[\sum_{x\in \Gamma_{1}, y\in \Gamma_{1}}L_{x \rightarrow y}+(n-1)\sum_{x\in \Gamma_{1}, y\in \Gamma_{2}}L_{x \rightarrow y}\right]\nonumber\\
    &=&nL_{\mathrm{total}}(t-1)+n(n-1)\sum_{x\in \Gamma_{1}, y\in \Gamma_{2}}(L_{x \rightarrow H_1}+1+L_{H_2 \rightarrow y})\nonumber\\
    &=&nL_{\mathrm{total}}(t-1)+n(n-1)N_{t-1}[2L_B(t-1)+N_{t-1}]\nonumber\\
    &=&nL_{\mathrm{total}}(t-1)+n^{2t-2}(n-1)[(n-1) 2^t-n+2],
 \end{eqnarray}
 and  Eq.~(\ref{Rec_LTotal}) is obtained.

 \section{Derivation of Eq.~(\ref{Rel_R_L})}
\label{sec: Rel_R_L}
Here we want to show that  Eq.~(\ref{Rel_R_L}) holds for  any two nodes $x$ any $y$ on $G_n(t)$ ($t\geq 1$).

For the case of $t=1$, the network is a clique $K_n$ with  $n$ nodes and  all the $n$ nodes  are equivalent to each other.
If we label the $n$ nodes as $1, 2, \cdots, n$, we have $T_{i \rightarrow j}=T_{1 \rightarrow 2}$ for any $i\neq j$, and
\begin{eqnarray}
T_{1 \rightarrow 2}&=&\frac{1}{n-1}+\frac{1}{n-1}[1+\sum_{i=3}^nT_{i \rightarrow 2}] \nonumber\\
      &=&1+\frac{n-2}{n-1}T_{1 \rightarrow 2}.
\label{EQFPTG1}
\end{eqnarray}
By solving the above equation, we get $T_{1 \rightarrow 2}=n-1$. Note that $E(1)=\frac{1}{2}n(n-1)$ and $$2E(1)\mathfrak{R}_{1\rightarrow 2}=T_{1 \rightarrow 2}+T_{2 \rightarrow 1}=2(n-1).$$
We find  $\mathfrak{R}_{1\rightarrow2}=\frac{2}{n}.$  However  $L_{1 \rightarrow 2}=1$. Therefore
 $\mathfrak{R}_{1\rightarrow2}=\frac{2}{n}L_{1 \rightarrow 2}$. And Eq.~(\ref{Rel_R_L})
holds for  any two nodes $x$ any $y$ of $G_n(1)$ because all the $n$ nodes  are equivalent to each other.

For the case of $t>1$,  clique $K_n$ is the basic unit for constructing the network $G_n(t)$. For any two nodes $x$ any $y$ of $G_n(t)$, they are connected through  several clique $K_n$. It is easy to verify Eq.~(\ref{Rel_R_L}) by using mathematical induction.


\section{Derivation of Eq.~(\ref{EQ_Sigma})}
\label{Sec:Sigma_t}
For $t=1$, $G_n(t)$ is  a clique $K_n$. ${L_{x \rightarrow y}}=1$ for any $x\in G_n(t), y\in G_n(t)$, and  $\Sigma_1=\frac{n-1}{n}$, which shows  Eq.~(\ref{EQ_Sigma}) holds for $t=1$.

For $t>1$, let $d_u(t)$ represents the degree of node $u$ in $G_n(t)$ and $d_u(t-1)$ represents the degree of node $u$ in subunit $\Gamma_i$, which is a copy of $G_n(t-1)$.   Due to the self-similar structure as shown in Fig.~\ref{fig:1}, we find
\begin{eqnarray}
&&\sum_{x, y \in \Gamma_1}\frac{d_x(t)}{2E(t)}\frac{d_y(t)}{2E(t)}{L_{x \rightarrow y}}\nonumber\\
&=&\left(\frac{E{(t-1)}}{E(t)}\right)^2\sum_{x, y \in \Gamma_1}\frac{d_x(t-1)}{2E{(t-1)}}\frac{d_y(t-1)}{2E{(t-1)}}{L_{x \rightarrow y}}\nonumber\\
   &&+\sum_{x=H_1, y \in \Gamma_1}\frac{[d_{H_1}(t)-d_{H_1}(t-1)]{d_y(t)}}{(2E(t))^2}{L_{x \rightarrow y}}\nonumber\\
   &&+\sum_{y=H_1, x \in \Gamma_1}\frac{[d_{H_1}(t)-d_{H_1}(t-1)]{d_x(t)}}{(2E(t))^2}{L_{x \rightarrow y}}\nonumber\\
   &=&\left(\frac{E{(t-1)}}{E(t)}\right)^2\Sigma(t-1)+\frac{(n-1)E{(t-1)}}{(E(t))^2}W_{B}(t-1),\nonumber
\label{Rec_Sigma_P1}
\end{eqnarray}
The last line of the above equation is obtained by noticing the fact that $H_1$ of $G_n(t)$ is also a peripheral node of $\Gamma_1$,  which is a copy of $G_n(t-1)$.

Similarly, for any $i\neq 1$,
\begin{eqnarray}
&&\sum_{x  \in \Gamma_1, y \in \Gamma_i}\frac{d_x(t)}{2E(t)}\frac{d_y(t)}{2E(t)}{L_{x \rightarrow y}}\nonumber\\
&=&\sum_{x  \in \Gamma_1, y \in \Gamma_i}\frac{d_x(t)}{2E(t)}\frac{d_y(t)}{2E(t)}(L_{x \rightarrow H_1}+1+L_{H_i \rightarrow y})\nonumber\\
&=&\frac{1}{n}\sum_{x  \in \Gamma_1}\frac{d_x(t)}{2E(t)}L_{x \rightarrow H_1}+\frac{1}{n^2}+\frac{1}{n}\sum_{y  \in \Gamma_i}\frac{d_y(t)}{2E(t)}L_{H_i \rightarrow y}\nonumber\\
&=&\frac{2}{n}\frac{E{(t-1)}}{E(t)}\sum_{x  \in \Gamma_1}\frac{d_x(t)}{2E(t)}L_{x \rightarrow H_1}+\frac{1}{n^2}\nonumber\\
   &=&\frac{2}{n}\frac{n^{t-1}-1}{n^t-1}W_{B}(t-1)+\frac{1}{n^2}.
\label{Rec_Sigma_P2}
\end{eqnarray}

Therefore
\begin{eqnarray}
\Sigma(t)&=&\sum_{x, y \in G_n(t)}\frac{d_x(t)}{2E(t)}\frac{d_y(t)}{2E(t)}{L_{x \rightarrow y}}\nonumber\\
   &=&n\sum_{x, y \in \Gamma_1}\frac{d_x(t)}{2E(t)}\frac{d_y(t)}{2E(t)}{L_{x \rightarrow y}}\nonumber\\
   &&+n\sum_{i=2}^n\sum_{x\in \Gamma_1, y \in \Gamma_i}\frac{d_x(t)}{2E(t)}\frac{d_y(t)}{2E(t)}{L_{x \rightarrow y}}\nonumber\\
   &=&n\left(\frac{n^{t-1}-1}{n^t-1}\right)^2\Sigma(t-1)+\frac{n-1}{n}\nonumber\\
  &&-\frac {2(n-1)n^{t-1}}{(n^t-1)^2}[{n^{t}-(n-1)(2n)^{t-1}-1}].
\label{Rec_Sigma}
\end{eqnarray}
 Using Eq.~(\ref{Rec_Sigma}) recursively, we obtain
 \begin{eqnarray}\label{ERQ_Sigma}
    \Sigma(t)
    &=&n^2\left(\frac{n^{t-1}-1}{n^t-1}\right)^2\Sigma(t-2)\nonumber\\
    &&+\frac {(n-1)}{n(n^t-1)^2}[(n^{t}-1)^2+n(n^{t-1}-1)^2]\nonumber\\
    &&-\frac {2(n-1)n^{t-1}}{(n^t-1)^2}[{n^{t}-(n-1)(2n)^{t-1}-1}]\nonumber \\
    &&-\frac {2(n-1)n^{t-1}}{(n^t-1)^2}[{n^{t-1}-(n-1)(2n)^{t-2}-1}]\nonumber \\
    &=&\cdots\nonumber \\
    &=&n^{t-1}\left(\frac{n^{t-1}-1}{n^t-1}\right)^2\Sigma(1)+\Theta_t-\Delta_t,
 \end{eqnarray}
where
 \begin{eqnarray}\label{Thetat}
\Theta_t&=&\frac {(n-1)}{n(n^t-1)^2}\sum_{k=0}^{t-2}n^k(n^{t-k}-1)^2\nonumber\\
       &=&\frac {(n-1)}{n(n^t-1)^2}\sum_{k=0}^{t-2}(n^{2t-k}-2n^{t}+n^k )\nonumber\\
       &=&\frac {(n-1)}{n(n^t-1)^2}\left[\frac{n^{2t+1}-n^{t+2}}{n-1}-2tn^t+\frac{n^{t-1}-1}{n-1}\right]\nonumber\\
       &=&\frac {n^{2t}-n^{t+1}}{(n^t-1)^2}-\frac {2tn^{t-1}(n-1)}{(n^t-1)^2}+\frac{n^{t-1}-1}{n(n^t-1)^2}
 \end{eqnarray}
 and
 \begin{eqnarray}\label{Deltat}
\Delta_t&=&\frac {2(n-1)n^{t-1}}{(n^t-1)^2}\sum_{k=2}^t[{n^{k}-(n-1)(2n)^{k-1}-1}]\nonumber\\
       &=&\frac {2(n-1)n^{t-1}}{(n^t-1)^2}\left[\frac{n^{t+1}-n^2}{n-1}-(n-1)\frac{(2n)^t-2n}{2n-1}\right]\nonumber\\
       &&-(t-1)\frac {2(n-1)n^{t-1}}{(n^t-1)^2}
 \end{eqnarray}
 Replacing $\Sigma(1)$ with $\frac{n-1}{n}$, plugging the expressions for $\Theta_t$ and $\Delta_t$  shown as Eqs. (\ref{Thetat}) and (\ref{Deltat})) into  Eq.~(\ref{ERQ_Sigma}),  Eq.~(\ref{EQ_Sigma}) is obtained.

\section{Derivation of Eq.~(\ref{WB})}
\label{Sec:WB}

For $t=1$, $G_n(t)$ is  a clique $K_n$. ${L_{u \rightarrow B}}=1$ for any $u\neq B$, and  $W_B(1)=\frac{n-1}{n}$, which shows  Eq.~(\ref{WB}) holds for $t=1$.

For $t>1$, let $d_u(t)$ represents the degree of node $u$ in $G_n(t)$ and $d_u(t-1)$ represents the degree of node $u$ in subunit $\Gamma_i$, which is a copy of $G_n(t-1)$.  We find, for any $i$ $(i=1,2,\cdots,n)$, $\sum_{x \in \Gamma_i}\frac{d_x(t)}{2E(t)}=\frac{1}{n}$, and  for any  $u\in\Gamma_i$,
 \begin{equation}
 d_u(t)=\left\{\begin{array}{ll} 2d_u(t-1)=2(n-1) & u=H_i\\ d_u(t-1) & u\neq H_i \end{array}\right.
 \end{equation}
Due to the self-similar structure as shown in Fig.~\ref{fig:1}, we have
 \begin{eqnarray}
W_B(t)&=&\sum_{u \in G_n(t)}\frac{d_u(t)}{2E(t)}{L_{u \rightarrow B}}\nonumber\\
   &=&\sum_{u \in \Gamma_1}\frac{d_u(t)}{2E(t)}{L_{u \rightarrow B}}+(n-1)\sum_{u \in \Gamma_2}\frac{d_u(t)}{2E(t)}{L_{u \rightarrow B}}\nonumber\\
   &=&\frac{E{(t-1)}}{E(t)}\sum_{u \in \Gamma_1}\frac{d_u(t-1)}{2E{(t-1)}}{L_{u \rightarrow B}}+\frac{(n-1)}{2E(t)}L_{H_1\rightarrow B}\nonumber\\
    &&+(n-1)\sum_{u \in \Gamma_2}\frac{d_u(t)}{2E(t)}({L_{u \rightarrow H_2}}+1+L_{H_1 \rightarrow B})\nonumber\\
      &=&\frac{E{(t-1)}}{E(t)}W_B(t-1)+\frac{(n-1)(2^{t-1}-1)}{2E(t)}\nonumber\\
    &&+(n-1)\sum_{u \in \Gamma_2}\frac{d_u(t)}{2E(t)}({L_{u \rightarrow H_2}}+2^{t-1})\nonumber\\
    &=&\frac{E{(t-1)}}{E(t)}W_B(t-1)+\frac{(n-1)(2^{t-1}-1)}{2E(t)}\nonumber\\
    &&+(n-1)\frac{E{(t-1)}}{E(t)}W_B(t-1)+\frac{n-1}{n}2^{t-1}\nonumber\\
    &=&\frac{nE{(t-1)}}{E(t)}W_B(t-1)+\frac{(n-1)(2^{t-1}n^t-1)}{n^{t+1}-n}.
\label{Rec_WB}
\end{eqnarray}
 Using Eq.~(\ref{Rec_WB}) recursively, we obtain
\begin{eqnarray}
W_B(t)&=&\frac{n^2E_{t-2}}{E(t)}W_B(t-2) \nonumber\\
&&+\frac{(n-1)}{n^{t+1}-n}[(2^{t-1}+2^{t-2})n^t-(1+n)]\nonumber\\
&=&\cdots\nonumber\\
&=&\frac{n^{t-1}E_{1}}{E(t)}W_B(1) \nonumber\\
&&+\frac{(n-1)}{n^{t+1}-n}[(2+2^2+\cdots+2^{t-1})n^t]\nonumber\\
&&-\frac{(n-1)}{n^{t+1}-n}(1+n+\cdots+n^{t-1}).
\label{ER_WB}
\end{eqnarray}
Plugging the expressions of $E(1)$, $E(t)$ and $W_B(1)$ into Eq.~(\ref{ER_WB}), we obtain Eq.~(\ref{WB}).

\section{Derivation of Eq.~(\ref{WH})}
\label{Sec:WH}
For $t=1$,  ${L_{u \rightarrow H_1}}=1$ for any $u\neq H_1$, and  $W_H(1)=\frac{n-1}{n}$, which shows  Eq.~(\ref{WH}) holds for $t=1$.

For $t>1$, due to the self-similar structure as shown in Fig.~\ref{fig:1}, we have
 \begin{eqnarray}
W_{H_1}(t)&=&\sum_{u \in G_n(t)}\frac{d_u(t)}{2E(t)}{L_{u \rightarrow H_1}}\nonumber\\
   &=&\sum_{u \in \Gamma_1}\frac{d_u(t)}{2E(t)}{L_{u \rightarrow H_1}}+(n-1)\sum_{u \in \Gamma_2}\frac{d_u(t)}{2E(t)}{L_{u \rightarrow H_1}}\nonumber\\
   &=&\frac{E{(t-1)}}{E(t)}\sum_{u \in \Gamma_1}\frac{d_u(t-1)}{2E{(t-1)}}{L_{u \rightarrow H_1}}\nonumber\\
    &&+(n-1)\sum_{u \in \Gamma_2}\frac{d_u(t)}{2E(t)}({L_{u \rightarrow H_2}}+1)\nonumber\\
      &=&n\frac{E{(t-1)}}{E(t)}W_{B}(t-1)+\frac{n-1}{n}.
\label{Rec_WB}
\end{eqnarray}
Replacing $E(t-1)$ and $W_{B}(t-1)$ from Eqs.~ (\ref{Total_Edges}) and (\ref{WB}) in Eq.~(\ref{Rec_WB}), we obtain  Eq.~(\ref{WH}).
{
\section{Derivation of Eq.~(\ref{Max_Min_Wy})}
\label{Sec:Comp}
For any given node $y \in G_n(t)$, we first prove that $W_{H_1}(t)\leq W_{y}(t)$, and then we prove that $W_{B}(t)\geq W_{y}(t)$.

Recalling the equivalence of the $n$ subunits $\Gamma_{i}$ ($i=1,2, ..., n$) and the equivalence for the $n$ central nodes $H_{i}$ ($i=1,2, ..., n$), without loss generality, we assume $y \in \Gamma_1$ and we refer to the distance from $y$ to $H_1$ as $L_{y\rightarrow H_1}=k$ $(k\geq 0)$.
For any node $u$ of $G_n(t)$, if $u \in \Gamma_1$,
  \begin{equation}
L_{u\rightarrow y}\geq L_{u\rightarrow H_1}-k,
 \end{equation}
while, if $u \not\in \Gamma_1$,
  \begin{equation}
L_{u\rightarrow y}= L_{u\rightarrow H_1}+k.
 \end{equation}
 Therefore,
 \begin{eqnarray}
W_{y}(t)&=&\sum_{u \in G_n(t)}\frac{d_u(t)}{2E(t)}{L_{u \rightarrow y}}\nonumber\\
   &=&\sum_{u \in \Gamma_1}\frac{d_u(t)}{2E(t)}{L_{u \rightarrow y}}+\sum_{i=2}^n\sum_{u \in \Gamma_i}\frac{d_u(t)}{2E(t)}{L_{u \rightarrow y}}\nonumber\\
   &\geq&\sum_{u \in \Gamma_1}\frac{d_u}{2E{(t)}}{(L_{u\rightarrow H_1}-k)}\nonumber\\
    &&+(n-1)\sum_{u \in \Gamma_2}\frac{d_u(t)}{2E(t)}({L_{u \rightarrow H_1}}+k)\nonumber\\
      &=&\sum_{u \in G_n(t)}\frac{d_u(t)}{2E(t)}{L_{u \rightarrow H_1}}+\frac{n-2}{n}k\nonumber\\
      &=& W_{H_1}(t)+\frac{n-2}{n}k       \nonumber\\
      &\geq & W_{H_1}(t).
\label{WH_Wy0}
\end{eqnarray}

 Now we come to prove  $W_{B}(t)\geq W_{y}(t)$  by induction.

 Base case: for $t=1$, all nodes are equivalent with each other and then $W_{B}(1)\geq W_{y}(1)$ for all $y$.

 Inductive step: we will show that $W_{B}(t)\geq W_{y}(t)$ holds if $W_{B}(t-1)\geq W_{y}(t-1)$ for  any $t>1$. Without loss of generality, we assume $y \in \Gamma_1$ and $B$ is one of the peripheral nodes in $\Gamma_1$. Therefore, $L_{H_1 \rightarrow y} \leq L_{H_1 \rightarrow B}$.  Note that $\Gamma_1$ is a copy of $G_n(t-1)$. We obtain
 \begin{eqnarray}
  \sum_{u \in \Gamma_1}\frac{d_u(t)}{2E(t)}{L_{u \rightarrow y}}\leq
  \sum_{u \in \Gamma_1}\frac{d_u}{2E{(t)}}{(L_{u\rightarrow B})}.
\label{WH_Wy}
\end{eqnarray}
 For any node $u \in G_n(t)$,
if $u \not\in \Gamma_1$,
 \begin{eqnarray}
L_{u\rightarrow y}&=& L_{u\rightarrow H_1}+L_{H_1 \rightarrow y}\nonumber\\
&\leq& L_{u\rightarrow H_1}+L_{H_1 \rightarrow B}=L_{u \rightarrow B}.
 \end{eqnarray}
 Therefore,
 \begin{eqnarray}
W_{y}(t)&=&\sum_{u \in G_n(t)}\frac{d_u(t)}{2E(t)}{L_{u \rightarrow y}}\nonumber\\
   &=&\sum_{u \in \Gamma_1}\frac{d_u(t)}{2E(t)}{L_{u \rightarrow y}}+\sum_{i=2}^n\sum_{u \in \Gamma_i}\frac{d_u(t)}{2E(t)}{L_{u \rightarrow y}}\nonumber\\
   &\leq&\sum_{u \in \Gamma_1}\frac{d_u(t)}{2E(t)}{L_{u \rightarrow B}}+\sum_{i=2}^n\sum_{u \in \Gamma_i}\frac{d_u(t)}{2E(t)}{L_{u \rightarrow B}}\nonumber\\
   &= & W_{B}(t).
\label{WH_Wy2}
\end{eqnarray}
Combing (\ref{WH_Wy0}) and (\ref{WH_Wy2}), we finally get
  \begin{equation}\label{Max_Min1_Wy}
 W_{H_1}(t)\leq W_{y}(t) \leq W_{B}(t).
 \end{equation}


\section{Derivation of Eq.~(\ref{eTSUBcenter})}
\label{sec: eTSUBcenter}

We first calculate $T_{H_2\rightarrow H_1}$ and $T_{H_1\rightarrow C_1}$, then we insert them into Eq.~(\ref{TSUBcenter}), in such a way that Eq.~(\ref{eTSUBcenter}) is obtained.

Let $x=H_2$ and $y=H_1$ in Eq.~(\ref{EMTTY}), one can get
\begin{eqnarray}
T_{H_2 \rightarrow H_1} &=&\frac{2E(t)}{n}[{L_{H_2 \rightarrow H_1}}+W_{H_1}(t)-W_{H_2}(t)]\nonumber\\
   &=&\frac{2E(t)}{n}{L_{H_2 \rightarrow H_1}}\nonumber\\
   &=&n^t-1.
\label{FH2H1}
\end{eqnarray}
In order to calculate  $T_{H_1\rightarrow C_1}$, we must derive $W_{C_1}(t)-W_{H_1}(t)$.
For any node $u\in G_n(t)$, if $u \not\in \Gamma_1$,
  \begin{equation}
L_{u\rightarrow C_1}= L_{u\rightarrow H_1}+ L_{H_1 \rightarrow C_1 };
 \end{equation}
 if $u \in \Gamma_1$ and $u \not\in \Gamma_{11}$,
  \begin{equation}
L_{u\rightarrow C_1}= L_{u\rightarrow H_1}- L_{H_1 \rightarrow C_1 };
 \end{equation}
  if $u \in \Gamma_{11}$,
 \begin{eqnarray}
  \sum_{u \in \Gamma_{11}}\frac{d_u(t)}{2E(t)}{L_{u \rightarrow H_1}}=
  \sum_{u \in \Gamma_{11}}\frac{d_u}{2E{(t)}}{L_{u\rightarrow C_1}}.
\end{eqnarray}
Thus,
 \begin{eqnarray}
W_{C_1}(t)&=&\sum_{u \in G_n(t)}\frac{d_u(t)}{2E(t)}{L_{u \rightarrow C_1}}\nonumber\\
   &=&\sum_{u \in \Gamma_{11}}\frac{d_u}{2E{(t)}}{L_{u\rightarrow C_1}}+\sum_{u \in \Gamma_1, u \not \in \Gamma_{11}}\frac{d_u(t)}{2E(t)}{L_{u \rightarrow  C_1}}\nonumber\\
   &&   +\sum_{i=2}^n\sum_{u \in \Gamma_i}\frac{d_u(t)}{2E(t)}{L_{u \rightarrow  C_1}}\nonumber\\
   &=&\sum_{u \in G_n(t)}\frac{d_u(t)}{2E(t)}{L_{u \rightarrow H_1}}-\sum_{u \in \Gamma_{11}}\frac{d_u(t)}{2E(t)}{L_{H_1 \rightarrow C_1}}\nonumber\\
   &&   +\sum_{i=2}^n\sum_{u \in \Gamma_i}\frac{d_u(t)}{2E(t)}{L_{H_1 \rightarrow C_1}}\nonumber\\
    &= & W_{H_1}(t)+\frac{n-1}{n}(1-\frac{E(t-1)}{E(t)})L_{H_1 \rightarrow C_1}\nonumber\\
   &= & W_{H_1}(t)+(2^{t-2}-1)(n-1)\frac{n^{t-1}-n^{t-2}}{n^{t}-1}.
\end{eqnarray}
Therefore,
 \begin{eqnarray}\label{WC_WHa}
  W_{C_1}(t)-W_{H_1}(t)=(2^{t-2}-1)(n-1)\frac{n^{t-1}-n^{t-2}}{n^{t}-1}.
 \end{eqnarray}

Let $x=H_1$ and $y=C_1$ in Eq.~(\ref{EMTTY}), and replace $W_{C_1}(t)-W_{H_1}(t)$ from Eq.~(\ref{WC_WHa}), we obtain
\begin{eqnarray}
T_{H_1 \rightarrow C_1} &=&\frac{2E(t)}{n}[{L_{H_1 \rightarrow C_1}}+W_{C_1}(t)-W_{H_1}(t)]\nonumber\\
   &=&(2^{t-2}-1)(2n^{t}-2n^{t-1}+n^{t-2}-1).
\label{FH1C1}
\end{eqnarray}

Replacing $T_{\Omega_0}(t-1)$, $T_B(t-1)$, $T_{H_2\rightarrow H_1}$ and $T_{H_1\rightarrow C_1}$ from Eqs.~(\ref{eTcenter}), (\ref{MTTB}), (\ref{FH2H1}) and (\ref{FH1C1}) in Eq.~(\ref{TSUBcenter}),  Eq.~(\ref{eTSUBcenter}) is obtained.

}

\end{document}